\documentclass[sigconf]{acmart}

\acmConference[ICSE 2024]{46th International Conference on Software Engineering}{April 2024}{Lisbon, Portugal}
\usepackage{adjustbox}
\usepackage{algorithmic}
\usepackage{amsmath,amsfonts}
\usepackage{balance}
\usepackage{caption}
\usepackage{comment}
\usepackage{graphicx}
\usepackage{hyperref}
\usepackage{listings,multicol}
\usepackage{multirow}
\usepackage{natbib}
\usepackage[normalem]{ulem}
\usepackage{textcomp}
\usepackage{xcolor}
\usepackage{soul}
\usepackage{wrapfig}
\copyrightyear{2024}
\acmYear{2024}
\setcopyright{rightsretained}
\acmConference[ICSE 2024]{2024 IEEE/ACM 46th International Conference on Software Engineering}{April 14--20, 2024}{Lisbon, Portugal}
\acmBooktitle{2024 IEEE/ACM 46th International Conference on Software Engineering (ICSE 2024), April 14--20, 2024, Lisbon, Portugal}\acmDOI{10.1145/3597503.3608141}
\acmISBN{979-8-4007-0217-4/24/04}
\def\BibTeX{{\rm B\kern-.05em{\sc i\kern-.025em b}\kern-.08em
    T\kern-.1667em\lower.7ex\hbox{E}\kern-.125emX}}

\begin{document}
\author{Adriana Sejfia}
\email{sejfia@usc.edu}
\affiliation{%
  \institution{University of Southern California}
  \state{California}
  \country{USA}
}

\author{Satyaki Das}
\email{satyakid@usc.edu}
\affiliation{%
  \institution{University of Southern California}
  \state{California}
  \country{USA}
}

\author{Saad Shafiq}
\email{saad4is@hotmail.com}
\affiliation{%
  \institution{Johannes Kepler University}
  \country{Austria}
}

\author{Nenad Medvidovi\'c}
\email{neno@usc.edu}
\affiliation{%
  \institution{University of Southern California}
  \state{California}
  \country{USA}
}

\newcommand{\adriana}[1]{{\color{blue}(Adriana: #1)}}
\newcommand{\neno}[1]{{\color{teal}(Neno: #1)}}

\newcommand{\lineVulTotalVulPatches}{3,286}

\newcommand{\lineVulFunctionComplexVul}{37}
\newcommand{\lineVulFunctionComplexPatches}{1,401}
\newcommand{\lineVulFunctionRepeatedVul}{189}
\newcommand{\lineVulFunctionComplexContribution}{75}

\newcommand{\lineVulLineRepeatedVul}{63}
\newcommand{\lineVulLineComplexPatches}{2,078}
\newcommand{\lineVulLineComplexVul}{61}
\newcommand{\lineVulLineComplexVulContribution}{96}

\newcommand{\lineVulFuncLevelTPRFunction}{0.87}
\newcommand{\lineVulFuncLevelTPRVul}{0.84}
\newcommand{\lineVulFuncLevelTPRAtomic}{0.88}
\newcommand{\lineVulFuncLevelTPRComplex}{0.62}

\definecolor{codegreen}{rgb}{0,0.6,0}
\definecolor{codegray}{rgb}{0.5,0.5,0.5}
\definecolor{codepurple}{rgb}{0.58,0,0.82}
\definecolor{backcolour}{rgb}{0.95,0.95,0.92}

\lstdefinestyle{patchstyle1}{
    commentstyle=\color{codegreen},
    keywordstyle=\color{magenta},
    numberstyle=\footnotesize\color{codegray},
    stringstyle=\color{codepurple},
    basicstyle=\ttfamily,
    breaklines=true,
    captionpos=b,
    keepspaces=true,
    numbersep=5pt,
    showspaces=false,
    showstringspaces=false,
    showtabs=false,
    tabsize=2
}

\title{Toward Improved Deep Learning-based Vulnerability Detection}
\begin{abstract}
Deep learning (DL) has been a common thread across several recent techniques for vulnerability detection. The rise of large, publicly available datasets of vulnerabilities has fueled the learning process underpinning these techniques. While these datasets help the DL-based vulnerability detectors, they also constrain these detectors' predictive abilities. Vulnerabilities in these datasets have to be represented in a certain way, e.g., code lines, functions, or program slices within which the vulnerabilities exist. We refer to this representation as a base unit. The detectors learn how base units can be vulnerable and then predict whether other base units are vulnerable. We have hypothesized that this focus on individual base units harms the ability of the detectors to properly detect those vulnerabilities that span multiple base units (or MBU vulnerabilities). For vulnerabilities such as these, a correct detection occurs when all comprising base units are detected as vulnerable. Verifying how existing techniques perform in detecting all parts of a vulnerability is important to establish their effectiveness for other downstream tasks. To evaluate our hypothesis, we conducted a study focusing on three prominent DL-based detectors: ReVeal, DeepWukong, and LineVul. Our study shows that all three detectors contain MBU vulnerabilities in their respective datasets. Further, we observed significant accuracy drops when detecting these types of vulnerabilities. We present our study and a framework that can be used to help DL-based detectors toward the proper inclusion of MBU vulnerabilities.

\end{abstract}
\maketitle

\section{introduction}
\label{sec:introduction}

 Software vulnerabilities have been the focus of a variety of studies. Researchers have tried to better understand vulnerabilities~\cite{santos2017, grobauer2011, bhatt2017, piessens2002}, discover new ways in which they can occur~\cite{liu2012},  automate their detection~\cite{pham2010, shin2010}, and so on. In particular, studies focusing on automated vulnerability detection have garnered increased interest. A common recent thread across these studies is their use of deep learning (DL) based techniques~\cite{li2021, zhou2019, cheng2021, chakraborty2021, li2018, hin2022, fu2022}, with \emph{publicly available datasets of vulnerable code} serving as drivers for this body of work.

Using these datasets, existing techniques attempt to learn how vulnerable code is manifested so that they can subsequently detect new occurrences. However, this is constrained by the representation of vulnerabilities in the datasets: some datasets highlight the \emph{functions}~\cite{chakraborty2021, zhou2019}, others the \emph{lines of code}~\cite{hin2022, fu2022}, and yet others the program dependence graph (PDG) \emph{slices}~\cite{cheng2021} that contain the vulnerabilities. In turn, the chosen representation serves as the \textit{base unit} for detection. Simply put, the existing DL-based detectors work by learning how instances of a given base unit can be vulnerable and then predicting whether other instances of the same base unit are vulnerable.

This strategy works well and is useful when all of the code pertaining to a vulnerability is contained within a single base unit. A concrete example of this type of vulnerability is CVE-2021-33815~\cite{cve202133815}, related to array access in the FFmpeg open-source library for A/V processing~\cite{ffmpeg}. This vulnerability existed in a single line, within a single function~{\cite{fix1cve20143647}}. A simplified view of the vulnerability and its fix is presented in Listing~\ref{lst:atomicexample}, with a leading ``+'' denoting added code and a leading ``--'' deleted code. Specifically, the vulnerability was caused by the way the \texttt{if} condition was checking for the size of the array to be allocated in the later call to \texttt{memset}.

\begin{center}
\begin{tabular}{c}
\captionsetup{margin=8pt, justification=justified,singlelinecheck=false}
\begin{lstlisting}[basicstyle=\fontsize{7}{9}\selectfont\ttfamily
\linespread{0.1},language=C, caption={ Fix for CVE-2021-38315 in FFmpeg}]
unsigned long dest_len = dc_count * 2LL;
- if (dc_count > (6LL*td->xsize*td->ysize+63)/64) 
+ if (dc_count != (td->xsize>>3)*(td->ysize>>3)*3)
        return INVALIDDATA;

memset(td->data, 0, dest_len + PADDING);
\end{lstlisting}
\label{lst:atomicexample}
\end{tabular}
\end{center}

\begin{figure}[!t]
    \centering
    \includegraphics[width=8.5cm]{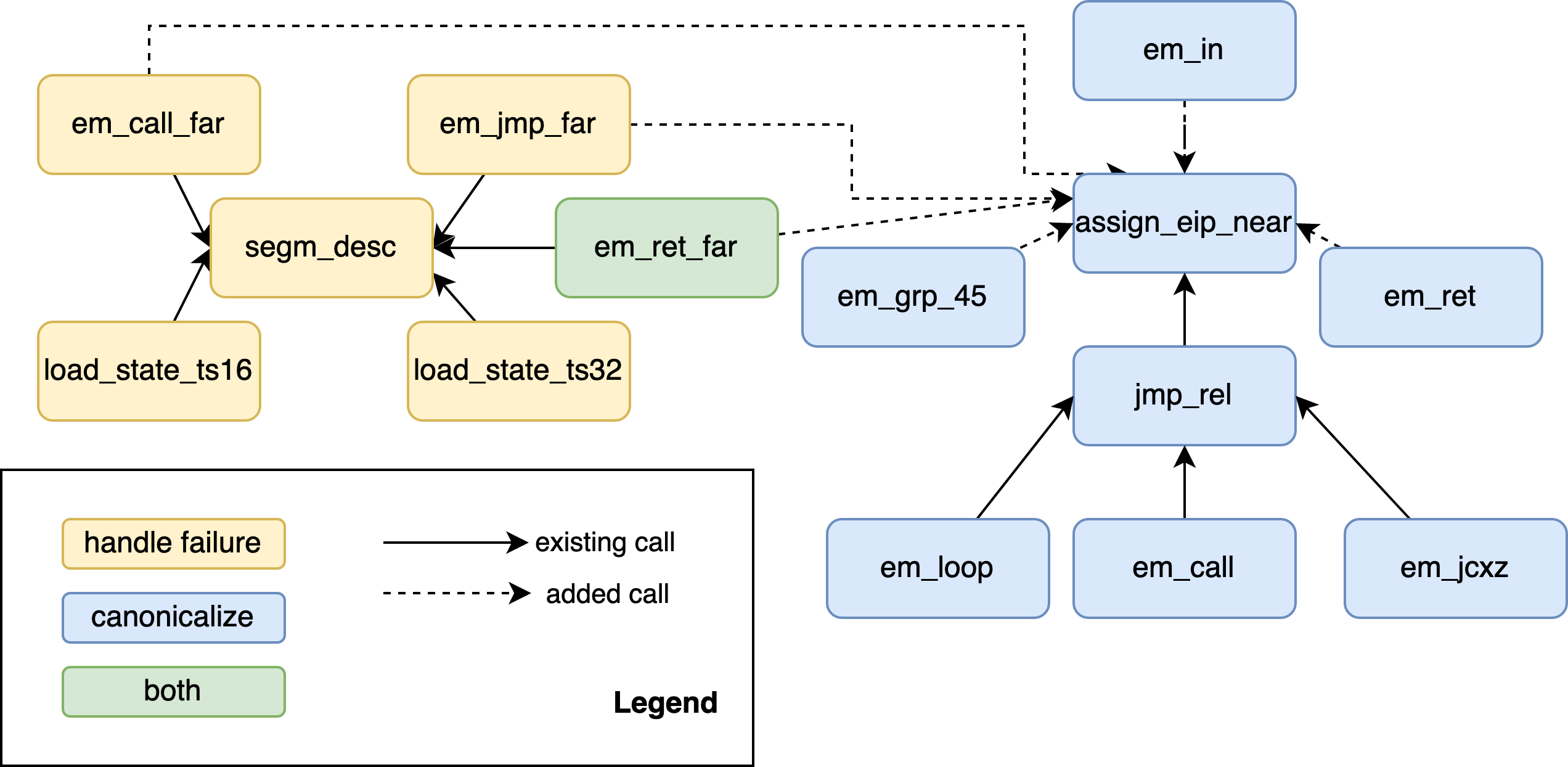}

    \caption{Some functions involved in the CVE-2014-3647 vulnerability. Each rectangle represents a function.}
    \label{fig:linux-complex-fig}
\end{figure}

While detecting such vulnerabilities is useful, \emph{many real-world vulnerabilities span more than one base unit} (line, function, or slice). An example is CVE-2014-3647~\cite{cve20143647}, discovered in the Linux kernel. This vulnerability was rooted in how the kernel changed the register (RIP) that contained the instruction to be executed when performing certain operations, such as \texttt{jmp}, \texttt{call}, or \texttt{ret}; the vulnerability ultimately could allow OS guest users to cause a denial of service. This vulnerability was fixed through two patches~\cite{fix1cve20143647, fix2cve20143647}. For illustration, a simplified view of the functions involved in the vulnerability, along with their interdependencies, is presented in Figure~\ref{fig:linux-complex-fig}. Specifically, the vulnerability was spread across (1)~functions that performed the above-mentioned \texttt{jmp}, \texttt{call}, and \texttt{ret} operations; (2)~functions that ensured the kernel is not left in an inconsistent state in cases of failure (\texttt{segm\_desc}); and (3)~functions that assigned the RIPs 
 (\texttt{assign\_eip\_near}). All these functions took part in the vulnerability, meaning that all of them (as well as several others not shown in the figure) had to be tagged as vulnerable in order for the vulnerability itself to be detected.

This discrepancy between the prevalent focus of DL-based vulnerability detectors on {individual base units} (IBUs) and the spread of many real-world vulnerabilities across {multiple base units} (MBUs) motivated us to \emph{examine how existing detectors perform on MBUs}. To our knowledge, no prior research has studied this problem. 
Specifically, for developers to understand a vulnerability that spans more than one base unit, all involved base units need to be detected as vulnerable. In the example of CVE-2014-3647 from Figure~\ref{fig:linux-complex-fig}, detecting only one involved line or function would not result in the successful mitigation of this vulnerability, still leaving the security threat in the code. Establishing the effectiveness of DL-based detectors for MBU vulnerabilities is also important for subsequent research. For example, a study focused on automated vulnerability patch generation may rely on a DL-based detector to automatically collect data on what code is vulnerable. 
Inaccurate detection of MBUs will directly harm such a study and all of its follow-ons. 

This is why we decided to examine how state-of-the-art DL-based detectors perform on complete vulnerabilities, as opposed to only their comprising IBUs. Our guiding hypothesis was that \emph{the performance of these detectors on MBU vulnerabilities is dependent on how they were trained}. This hypothesis was inspired by several pilot experiments we ran on DL-based detectors with publicly available models. Specifically, we obtained results of each detector's performance on data used in studies reported for other detectors. In each case, we saw a significant drop in a detector's accuracy as compared to its published results.

As an example, in one of these experiments, we considered two prominent DL-based detectors: ReVeal~\cite{chakraborty2021}, a function-level pipeline for vulnerability detection, and DeepWukong~\cite{cheng2021}, a slice-level detector. As part of the experiments, we checked the performance of DeepWukong on ReVeal's dataset. We first obtained slices from ReVeal's functions and then ran DeepWukong's model on the transformed dataset. DeepWukong's reported accuracy on its own dataset is always above 90\%, but when applied to the dataset used by ReVeal the accuracy dropped to 56\%. Such significant decreases recurred across our pilots, 
whenever an existing detector was applied on data drawn from different distributions than those on which it has been trained. This led us to postulate that, \emph{if these
state-of-the-art detectors did not explicitly take into account vulnerabilities that span more than one of their base unit of choice, their performance on MBU vulnerabilities will analogously suffer}.

For our in-depth study of this issue, we focused on three prominent DL-based vulnerability detectors: the above-mentioned ReVeal and DeepWukong, as well as the line-level detector LineVul~\cite{fu2022}.
We scrutinized the data, approaches, and implementations of the three detectors to answer the following research questions: 
    \begin{itemize}
        \item \textbf{RQ1:} \emph{What percentage of the vulnerabilities in the detectors' datasets are MBU vulnerabilities?} 
        Our hypothesis was that if MBU vulnerabilities formed only a negligible part of the datasets, these datasets were not representative of real-world vulnerabilities. However, we discovered that MBU vulnerabilities' presence in these datasets is significant: they comprise 22\% of all vulnerabilities in ReVeal's dataset, 53\% in  DeepWukong's dataset, and \lineVulLineComplexVul\% and \lineVulFunctionComplexVul{}\%, respectively, in two different configurations of LineVul's dataset. The complete results can be found in Section~\ref{sec:RQ1}.
        \item \textbf{RQ2:} \emph{How are the comprising parts of MBU vulnerabilities used in the training and evaluation of the detectors?} 
        Our goal was to determine whether the detectors consider realistic scenarios by grouping all base units pertaining to a single vulnerability when training, validating, testing, and reporting the accuracies. We found that the three detectors fail to properly take MBU vulnerabilities into account. More details on the findings can be found in Section~\ref{sec:RQ2}. The answers to this question also spawned RQ3 and RQ4.
        \item \textbf{RQ3:} \emph{How accurate are the detectors in actually uncovering vulnerabilities?} Since the accuracy reports of the vulnerability detectors focus on base units as opposed to vulnerabilities, 
        we analyzed the three detectors' evaluation results to obtain their accuracies on complete vulnerabilities, by grouping all base units of each MBU vulnerability prior to reporting. Since we were specifically looking at the vulnerable data in the dataset, we computed (1)~the true positive rates (TPR), (2)~precision, and (3)~Matthews correlation coefficient (MCC), which was included due to its applicability in cases of imbalanced datasets, a scenario which vulnerability detectors face.  We found that the values of the three metrics generally drop when considering complete vulnerabilities as opposed to their constituent base units. The details of our analysis can be found in Section~\ref{sec:RQ3}.
        \item \textbf{RQ4:} \emph{Does training and evaluating the vulnerability detectors in a more realistic way affect their accuracies?} For this question, we retrained the detectors by adjusting the division of the data into training, validating, and testing sets. When dividing the data across these three sets, we focused on \textit{complete vulnerabilities} instead of base units. In other words, we implemented a constraint that ensured a single MBU's base units are all assigned to one of the training, validating, or testing sets, and are not spread out amongst the three sets. We found that the three metrics' values are impacted by this, more realistic training and testing approach. Our findings are reported in Section~\ref{sec:RQ4}.
    \end{itemize}



Some of the prior work we studied did consider MBU vulnerabilities~\cite{chakraborty2021}, but mostly expressed concerns that such vulnerabilities may contain noise and would thus pollute the learning process. We manually analyzed a portion of the relevant datasets to verify these claims. Our analysis revealed that (1) this is not always the case and (2) when that does happen, there are ways to automatically identify true MBU vulnerabilities, as described in Section~\ref{sec:identification}. Given this finding and the difficulties existing detectors experience with MBU vulnerabilities (Section~\ref{sec:findings}), in Section~\ref{sec:framework} we present an automated framework to enable MBU vulnerabilities' appropriate inclusion in DL-based detectors. We make the artifacts from our study and the components of our framework publicly available~\cite{ourwebsite}.

The key contributions of this paper are thus three-fold:
\begin{itemize}
      \item {A \emph{definition and categorization} of MBU vulnerabilities.} 
      \item {A \emph{systematic analysis} of the prevalence and detection accuracy of MBU vulnerabilities in state-of-the-art DL-based detectors.} 
      \item {An \emph{automated framework} for appropriate inclusion of MBU vulnerabilities in DL-based detection}. 
    \end{itemize}

\section{MBU vulnerabilities}
\label{sec:definition}

The comprising base units (program lines, slices, or functions) of an MBU vulnerability combine to contribute to the vulnerability. While each of these base units may contribute to different aspects of the vulnerability, it is through their interplay that the vulnerability is manifested. Going back to the example of  CVE-2014-3647~\cite{cve20143647} (recall Figure~\ref{fig:linux-complex-fig}), this vulnerability involved several functions that invoked each other. Two primary issues that, in tandem, created this vulnerability were (1)~not checking whether certain destinations were canonical (functions colored in blue and green) before making RIP changes and (2)~improperly handling faults in RIP loading (functions colored in yellow and green). 



\looseness=-1
The characterization of MBU vulnerabilities is important because it also helps us identify vulnerabilities that are \textit{not} MBU. Before we dwell on this, a bit of additional context is necessary. The vulnerability datasets used in DL-based detectors are primarily collected by focusing on the code patches that fixed the vulnerabilities: by considering the parts of the code that were changed to fix a vulnerability, one can derive which code locations were vulnerable in the first place. For instance, in datasets where the base unit is a function, all functions that were changed in the patch(es) that fixed a given vulnerability are collected (and sometimes cleaned in order to reduce the presence of irrelevant changes) and added to the dataset. 

One observation we made based on a manual analysis of the existing datasets is that while patches themselves may be \textit{compound}, i.e., they introduce changes to multiple base units, this does not necessarily mean those patches fix MBU vulnerabilities.
Specifically, compound patches may fix a vulnerability that is \textit{independently} manifested in multiple code locations, that is, the code locations are similarly vulnerable and they are not dependent on each other. In cases when an error was repeatedly made in the code, causing the same vulnerability in various locations, developers often opt to fix many or all of these manifestations of the same vulnerability in the same vulnerability-fixing patch. 
 \begin{adjustbox}{center}
  \begin{minipage}{\columnwidth}  
\begin{lstlisting}[basicstyle=\fontsize{7}{8}\selectfont\ttfamily\linespread{0.7},language=C, caption=Buffer overflow issue in QEMU]
                FunctionA(){
                ...
                -   if (offset > 0x200)
                +   if (offset >= 0x200) 
                ...
                }
                FunctionB(){
                ...
                -   if (offset > 0x200)
                +   if (offset >= 0x200) 
                ...
                }
                FunctionC(){
                ...
                -   if (offset > 0x200)
                +   if (offset >= 0x200) 
                ...
                }
                 ....
\end{lstlisting}
\label{lst:repeated-atomic}
\end{minipage}
\end{adjustbox}


\looseness=-1
As an example, this happened with a recurring buffer overflow vulnerability in the open-source emulator QEMU~\cite{qemu}. A partial rendition of the issue and its corresponding fix can be found in Listing~\ref{lst:repeated-atomic}, following the same formatting convention as presented above~\cite{fixqemubufferoverflow}. The issue, present in the deleted lines, arose from the fact a buffer's length (\texttt{offset}) was not checked properly before writing data. 

The fixing patch modified the corresponding checks \emph{in the same way} in the three functions present in the figure, as well as three additional functions that have been elided for space. The vulnerability in each of these functions is independent of the other functions that were analogously changed. 
 We term these types of vulnerabilities \textit{repeated IBU}.\footnote{Note that, as with MBU vulnerabilities, the definition of what constitutes a repeated IBU vulnerability is dependent on the base unit. For example, a vulnerability may be repeated IBU when the base unit is a function, but not when it is a line, if the changes in the function are spread across multiple lines.} We want to note that repeated IBU vulnerabilities are not duplicate vulnerabilities. While the underlying issue is the same in the many code locations they exist, the context in which they exist and the way in which they can be misused may differ.


Vulnerability patches may contain irrelevant changes to the fix of the vulnerability. Previous work has pointed to the fact that security fixes can induce other logic-preserving changes, referred to as casualty~\cite{sejfia2021} and trivial~\cite{kawrykow2011} changes. While necessary to make the code work, these changes do not reveal information about how that code is vulnerable, and they may end up inflating the size of a patch and obscuring the source of the vulnerability.  That is why identifying whether a compound patch fixes an MBU vulnerability involves isolating the irrelevant changes. There are automated tools that help towards that goal~\cite{sejfia2021, kawrykow2011}. Sometimes, collecting the vulnerability data involves manually labeling the vulnerable status of individual base units~\cite{zhou2019}, reducing the likelihood that irrelevant changes are present in the patches. 

\section{Distinguishing compound patches}
\label{sec:identification}

Before we started our study of MBU vulnerabilities, we needed to correctly identify them. From the datasets used in the three detectors, we were able to only obtain information about which vulnerability-fixing patches were compound but not which vulnerabilities were MBUs. As discussed in the previous section, compound patches can fix either MBU vulnerabilities or repeated IBU vulnerabilities. The challenge we needed to address was to correctly distinguish the patches that fixed MBU vulnerabilities from those that fixed repeated IBU vulnerabilities. This would permit us to identify MBU vulnerabilities and conduct the rest of the study. 

The problem of distinguishing compound patches was essentially one of establishing how similar changes in a patch are. Repeated IBU vulnerabilities contain highly similar, repetitive changes, as can be seen in the example of the buffer overflow vulnerability depicted in Listing~\ref{lst:repeated-atomic}. The same change happened in multiple locations in the code. MBU vulnerabilities follow more intricate patterns of code changes. This is what happened with the fix for the CVE-2017-0596 vulnerability in Android~\cite{cve20170596}. This was an elevation of privilege vulnerability and malicious actors could execute arbitrary code remotely. The fix involved making changes to various functions, three of which are shown in Listing~\ref{lst:complex-android}. The changes in these three functions vary significantly, especially when compared to the changes that fixed the vulnerability in Listing~\ref{lst:repeated-atomic}. In the first function, if conditions were modified and calls and assignments were deleted. The second function did not exist before and in the third function, a call to this second function was added.  

Beyond establishing the similarity of the changes, we needed an approach that could tell us if all of the changes in a patch were \textit{sufficiently} similar to belong to one group or not. If the changes belong to one group, then we can conclude a compound patch had repeated and very similar changes, and thus, fixes a repeated IBU vulnerability. Otherwise, we can conclude the patch fixes an MBU vulnerability. Clustering is one approach that helps with grouping elements based on their similarity. That is why we considered reusing approaches employed to cluster code changes, such as the C3 approach proposed by Kreutzer et al.~\cite{kreutzer2016}. 

C3 takes a history of commits from a repository, groups changes based on functions, computes the pairwise similarity of such changes, and finally clusters them. C3 can be configured to (1) represent the changes via lines of code or Abstract Syntax Trees (ASTs), and (2) use a hierarchical clustering algorithm (HCA) or the DBSCAN algorithm for the clustering part. These two clustering algorithms work well in cases when one does not know a priori the number of clusters to expect.

The C3 approach by and large applies to the problem of distinguishing compound patches with two exceptions. First, for our purposes, one only needs to consider the commit/s (i.e., the patches) that fixed a given vulnerability and not the whole history of commits. This is a simplification of C3's original approach as we would not need to conduct pairwise comparisons for changes that have happened throughout the history of a system. Second, because the definition of compound patches is determined by the base unit, our grouping of the changes also had to be done using the various base units, and not just functions. To fit the scenario of compound patches, we modified C3's approach to address these two exceptions. Since C3 does not have a publicly available implementation, we implemented it along with the two modifications ourselves. Further, in our implementation of C3, we picked representing code changes using ASTs and to use DBSCAN for the clustering portion; in C3's empirical results, ASTs and DBSCAN were shown to perform better than the other two options (line-based changes and HCA, respectively). Note that our two modifications do not alter the core algorithm of C3, just when and how it is applied. 

\begin{adjustbox}{center}
  \begin{minipage}{\columnwidth}
  \captionsetup{justification=raggedleft, singlelinecheck=false, margin=22pt}
\begin{lstlisting}[basicstyle=\fontsize{7}{8}\selectfont\ttfamily\linespread{0.8},language=C, caption=Fix for CVE-2017-0596 in Android]
        ERRORTYPE SoftMPEG4Encoder::releaseEncoder() {
        -    if (!mStarted) {
        -        return OMX_ErrorNone;
        +    if (mEncParams) {
        +        delete mEncParams;
        +        mEncParams = NULL;
             }
        -
        -    mStarted = false;
        +    if (mHandle) {
        +        delete mHandle;
        +        mHandle = NULL;
        +    
        }
        
        +void SoftAACEncoder2::onReset() {
        +    delete[] mInputFrame;
        +    mInputFrame = NULL;
        +    mInputSize = 0;
        +
        +    mSentCodecSpecificData = false;
        +    mInputTimeUs = -1ll;
        +    mSawInputEOS = false;
        +    mSignalledError = false;
        +}
        
         SoftAACEncoder2::~SoftAACEncoder2() {
        -    delete[] mInputFrame;
        -    mInputFrame = NULL;
        +    onReset();
         }
         
\end{lstlisting}
\label{lst:complex-android}
\end{minipage}
\end{adjustbox}

C3 was originally evaluated on general patches. Since we specifically wanted to use this approach for MBU vulnerability patches, we collected a representative ground truth to measure its accuracy in this context. Our ground truth was obtained from a sample of the available dataset of ReVeal, i.e., the FFmpeg~\cite{ffmpeg} and QEMU~\cite{qemu} security commits. Both of these systems are large, have been used in practice for more than two decades, and have a rich history of publicly available vulnerabilities. Because of their history, we were confident that the compound patches they contain are representative of vulnerabilities in general. To obtain a sample size, we used the well-known margin of error sample size formula and followed common practice for the parameters of the formula: 90\% for the confidence level and 10\% for the margin of error~\cite{singh2014}. The formula and the parameters resulted in a sample of 67 patches. Two software engineering researchers with several years of experience labeled the changes across these 67 patches and discussed disagreements. Disagreements that could not be resolved between the two engineers, were resolved by a third researcher with more than seven years of experience. Ultimately, out of the 67 patches, 19 were concluded to be repeated IBU, and the remaining 48, were MBU. 

The modified C3 approach yielded a precision of 84\%, recall of 78\%, and overall accuracy of 73\%, where our class of interest (positive class) was MBU vulnerabilities. Accuracy metrics around 75\% are generally acceptable in the community, but, we were concerned about missing 22\% of MBU vulnerabilities (as the recall metric was 78\%). Our subsequent analyses of establishing the usefulness of the detectors for MBU vulnerabilities would be negatively affected by it. In 22\% of the cases when we would have MBU vulnerabilities, we would consider them as repeated IBU. This would harm our conclusions. Additionally, we envision a future in which DL-based detectors, prior to training, identify which vulnerabilities in their dataset are MBU so that they can ensure these vulnerabilities are considered appropriately in their training, testing, and reported accuracies (we present more of this vision in Section~\ref{sec:framework}). If the C3 approach is used in the future to distinguish compound patches prior to the training of DL-based detectors, we argue having a high recall is important, even if a trade-off between precision and recall is needed.  This is so because, with a low recall, the detectors would learn that the MBU vulnerabilities that are not recalled are really repeated IBUs and should be broken down to their base units. This would harm the effectiveness of these detectors when used by developers or when used in subsequent studies. 

Seeking to boost the recall especially we went back to the ground truth data. Our initial observation regarding the similarity of the changes in repeated IBU vulnerabilities as compared to MBU vulnerabilities gave us an idea. We checked to see if using a simple similarity threshold would provide us with better results, especially better recall. To calculate the similarity, we reused C3's longest common subsequence. We empirically set a minimum similarity threshold between code changes belonging to different base units in a compound patch to 70\%. We (1) checked pairwise similarities between all code changes per base unit in a patch, (2) obtained the minimum out of all the similarities, and (3) if the minimum passed our threshold we concluded the vulnerability the patch fixed was repeated IBU; otherwise, the vulnerability was deemed to be MBU. The similarity-based approach resulted in a precision of 83\%, perfect recall, and overall accuracy of 85\%. Since the precision remained almost unchanged, and the recall and accuracy went up, the similarity-based approach was superior to the more convoluted clustering-based approach. 
However, since we derived the minimum similarity threshold from the same data we used for the accuracy calculation, to ensure we were not overfitting, we ran our similarity-based approach in another verification dataset. This verification dataset comprised 67 patches from Chromium, Android, php, Linux kernel, Qemu, and FFMPEG. An experienced researcher manually verified the results of the similarity-based approach. We make the manually obtained verification dataset publicly available on our website~\cite{ourwebsite}. The precision and recall in the verification dataset were both 95\%. This increased our confidence that the similarity-based approach can be used to correctly and accurately distinguish compound patches. 

\vspace{-0.1cm}

\vspace{-0.3cm}
\section{Findings}
\label{sec:findings}
\looseness=-1
With an approach that helped us identify MBU vulnerabilities, we set to answer our four research questions using the three DL-based detectors. These detectors' datasets had different particulars that needed to be considered. We will briefly go over these particulars before we explain the analyses we conducted as part of our study.

\subsection{Obtaining and curating datasets}
\label{sec:prerq}
In its dataset, ReVeal contains functions that are labeled as vulnerable and non-vulnerable. For our analysis, we needed access to changes that happened to the vulnerable functions. Since our approach to detecting MBU vulnerabilities relies on patches, we needed access to metadata, i.e., links or hashes, that could help us identify and retrieve the patch. ReVeal released patch hashes for only a portion of the functions in their dataset, the ones obtained in the FFmpeg and QEMU systems. We contacted the authors of ReVeal for patch information for the remaining of the data in their dataset but they informed us that due to an error in the data collection, they did not have it. We identified MBU patches from the portion of the data with publicly available information, removed IBU vulnerabilities to run our analysis, and present results only on that portion of the data. 

We had a similar problem with the dataset used in DeepWukong. DeepWukong's base unit is PDG slices. Their dataset also contains test cases that correspond to one vulnerability: each test case is used to generate vulnerable slices. The subject systems used in this dataset were the synthetically generated SARD~\cite{sard}, redis~\cite{redis}, and lua~\cite{lua}. The vulnerabilities of the last two systems were obtained through patches, but the dataset did not have information about those patches' metadata.  We reached out to DeepWukong's authors as well for this information but they informed us they no longer possess it. Our attempts to trace the available code in DeepWukong's dataset to certain commits in the repositories of their subject systems also failed: the code in the dataset is a modified version of the original code and cannot be traced. Because of this, we could not run our original check for repeated IBU vulnerabilities in DeepWukong's dataset. There was still an analysis we could perform with the data we had, however: if we assume that all test cases that generate multiple slices per vulnerability are in fact MBUs, this would give us the upper bound of DeepWukong's accuracy. We ran our analysis with this assumption. 

Finally, LineVul uses the BigVul~\cite{fan2020} dataset for its training and evaluation. The authors report LineVul's performance on both detecting vulnerable functions and localizing vulnerable lines. We checked the BigVul dataset for MBU vulnerabilities using both these base units (functions and lines). 

Using these three datasets and the respective DL-based approaches, in our study, we focused on four research questions. First, we looked into (1) quantifying the presence of MBU vulnerabilities in the datasets, and (2) establishing how the three approaches use MBU vulnerabilities in their training, validation, and testing sets, and how the MBU vulnerabilities are included in the reported accuracies. Then, because our results revealed MBU vulnerabilities are not properly included in training and evaluation, we also looked into (3) obtaining the accuracies of the three detectors on complete vulnerabilities, as opposed to individual base units, and finally (4) measuring the impact training and evaluating with complete vulnerabilities in mind would have on the accuracies of the three detectors. Note that while all three of the datasets also contain non-vulnerable samples, e.g., base units, or even entire patches, that were deemed to not fix vulnerabilities, we focused most of our analyses only on those that were labeled as vulnerable. In total, we analyzed 1,587 vulnerabilities used in ReVeal, 3,662 used in DeepWukong, and 2,078 used in LineVul. Next, we discuss the results we obtained from this study through the four research questions. 
 \begin{figure}[!b]
 \vspace{-1cm}
    \centering
    \includegraphics[width=8cm]{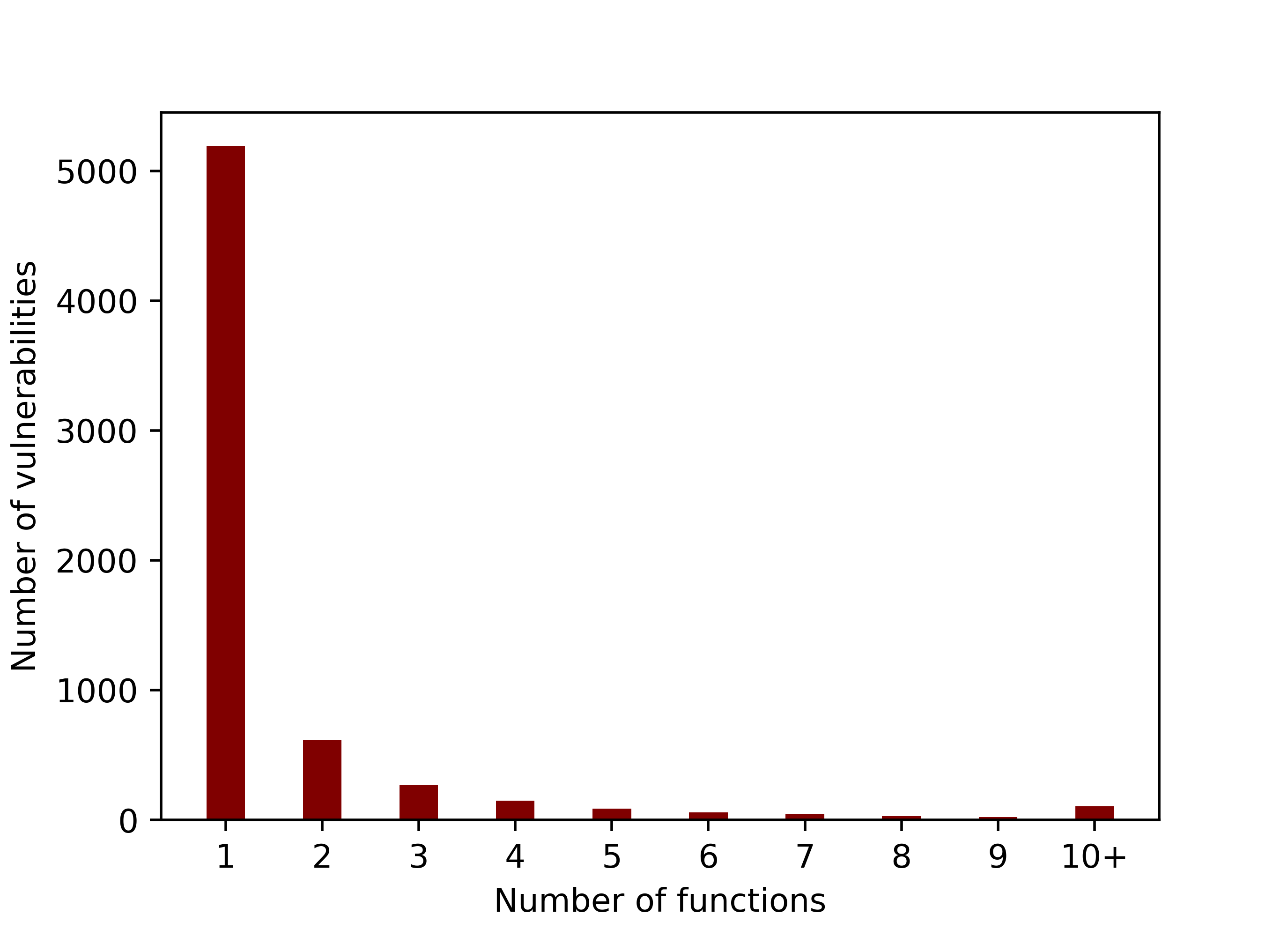}

    \caption{Vulnerabilities in the ReVeal dataset, grouped by their number of functions }
    \label{fig:reveal-func-dist}
\end{figure}
\subsection{RQ1: Presence of MBU vulnerabilities}
\label{sec:RQ1}
We present and discuss our results on the presence of MBU vulnerabilities across the three datasets.

1) \textit{ReVeal:} The vulnerable functions in the ReVeal dataset were obtained from 6,195 vulnerability-fixing patches.  1,587 of these patches were compound, which in the case of ReVeal means they changed more than one function. Out of the compound patches, our automated approach found 210 were repeated IBU vulnerabilities, meaning that the remaining 1,377 or 22\% of the vulnerabilities were MBUs. 55\% of the overall vulnerable functions in the ReVeal dataset came from the MBU vulnerabilities. We present the distribution of the number of functions in all of the ReVeal vulnerabilities in Figure~\ref{fig:reveal-func-dist}.

\begin{figure}[t!]
 \vspace{-0.5cm}
    \centering
    \includegraphics[width=8cm]{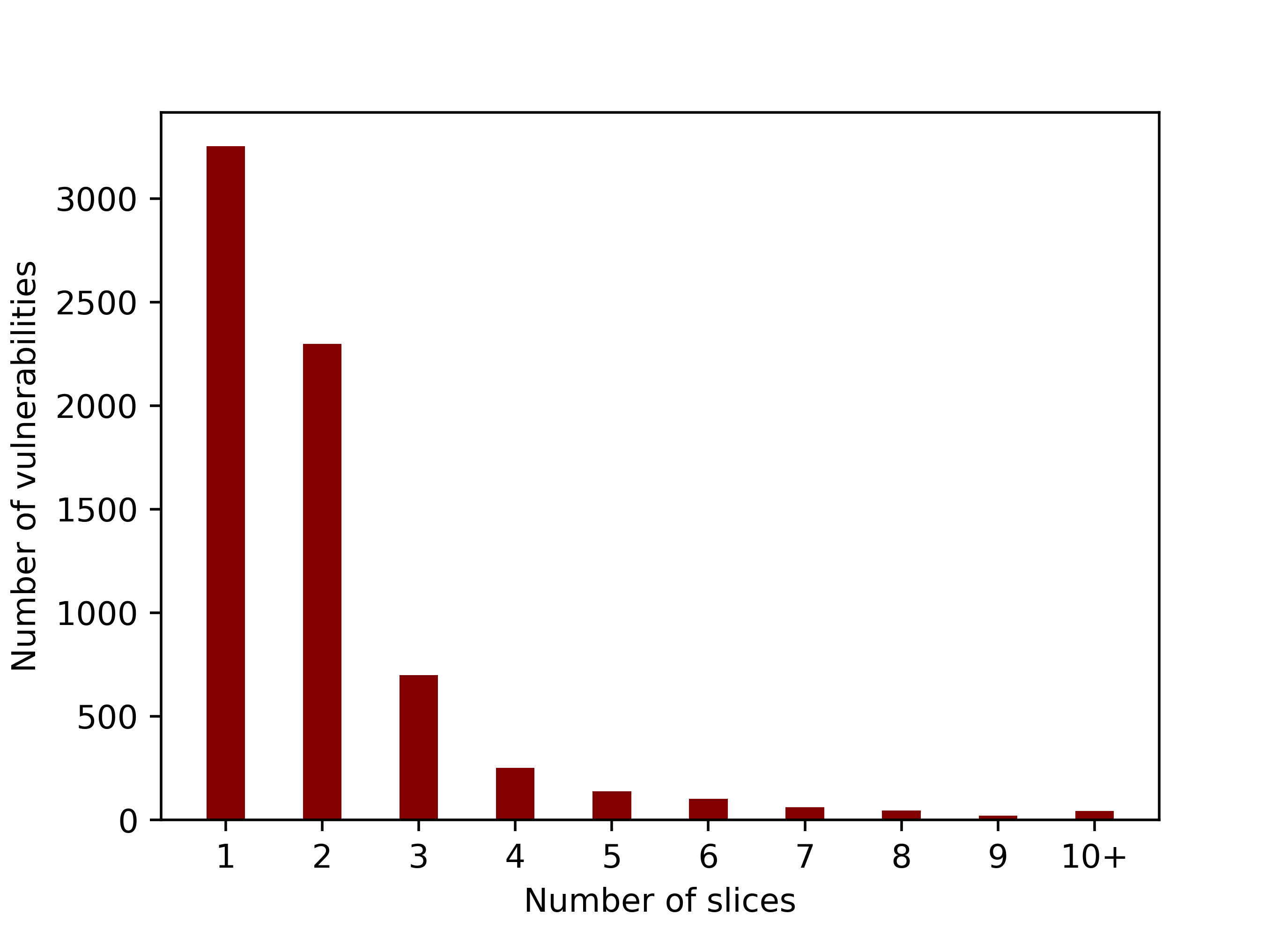}

    \caption{Vulnerabilities in the DeepWukong dataset, grouped by their number of slices }
    \label{fig:deepwukong-slices-dist}
    \vspace{-0.6cm}
\end{figure}

2) \textit{DeepWukong:} 53\% of the 6,911 vulnerabilities in DeepWukong span more than one PDG slice and thus are MBU vulnerabilities. The slices obtained from the MBU vulnerabilities account for 77\% of the overall vulnerable slices in this dataset. The distribution of slices across all the vulnerabilities in DeepWukong can be seen in Figure~\ref{fig:deepwukong-slices-dist}.

\begin{figure}[b!]
    \centering
    \includegraphics[width=8cm]{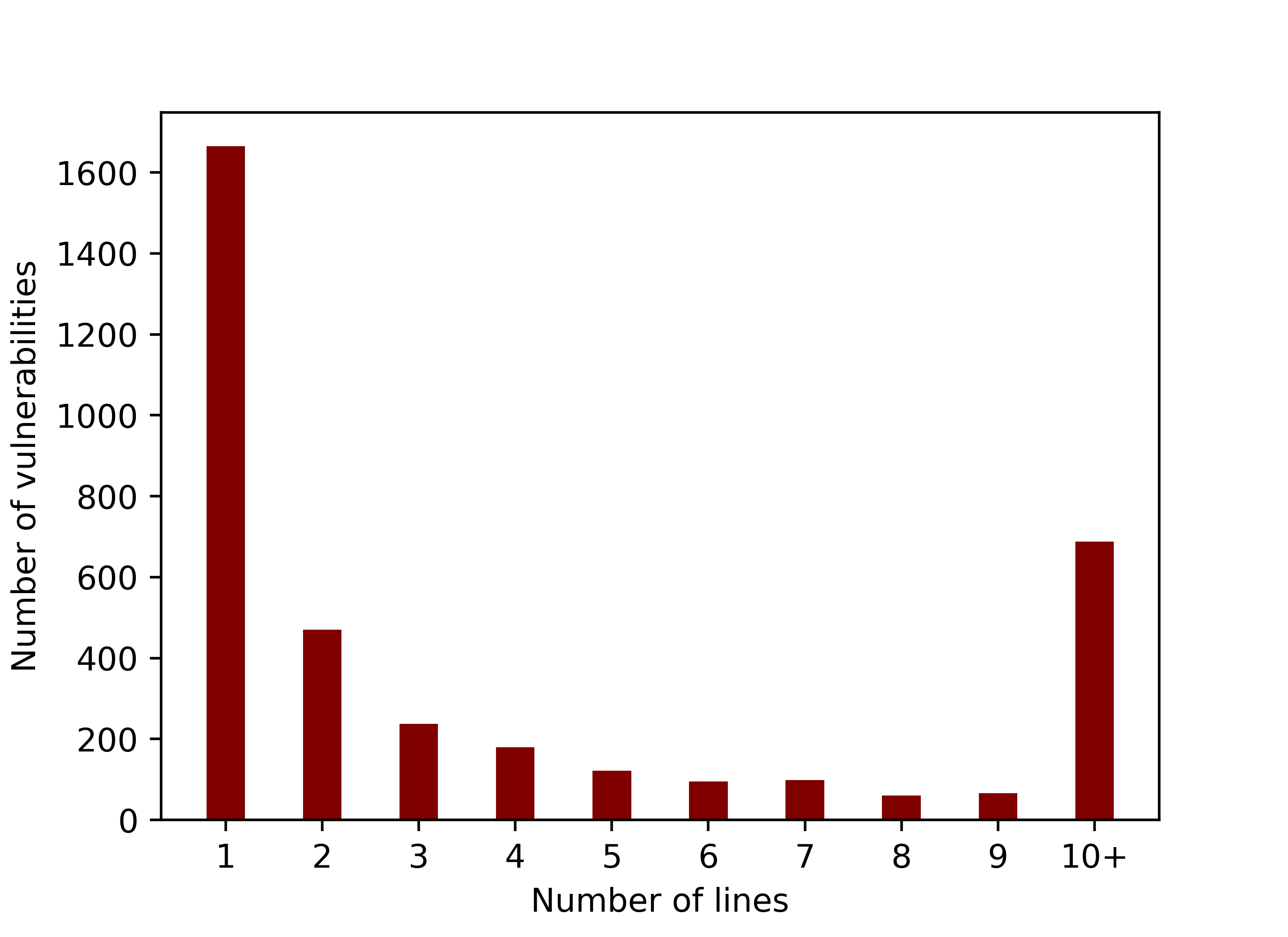}
    \caption{Vulnerabilities in the LineVul dataset, grouped by their number of lines }
    \label{fig:linevul-line-dist}
\end{figure}

3) \textit{LineVul:} We present two different views of the LineVul dataset: separated based on lines and based on functions. When looking at lines, out of a total of \lineVulTotalVulPatches{} vulnerabilities, \lineVulLineComplexPatches{} came from line-based compound patches. Our automated approach found that \lineVulLineRepeatedVul{} of those patches were actually fixing repeated IBU vulnerabilities. In total, out of all the vulnerabilities in LineVul's dataset, 2,015 or \lineVulLineComplexVul{}\% were line-based MBU vulnerabilities. The vulnerable lines in these MBU vulnerabilities accounted for \lineVulLineComplexVulContribution{}\% of all the vulnerable lines in the LineVul dataset. We present the distribution of the lines per vulnerability in Figure~\ref{fig:linevul-line-dist}.

\begin{figure}[t!]
    \centering
    \includegraphics[width=8cm]{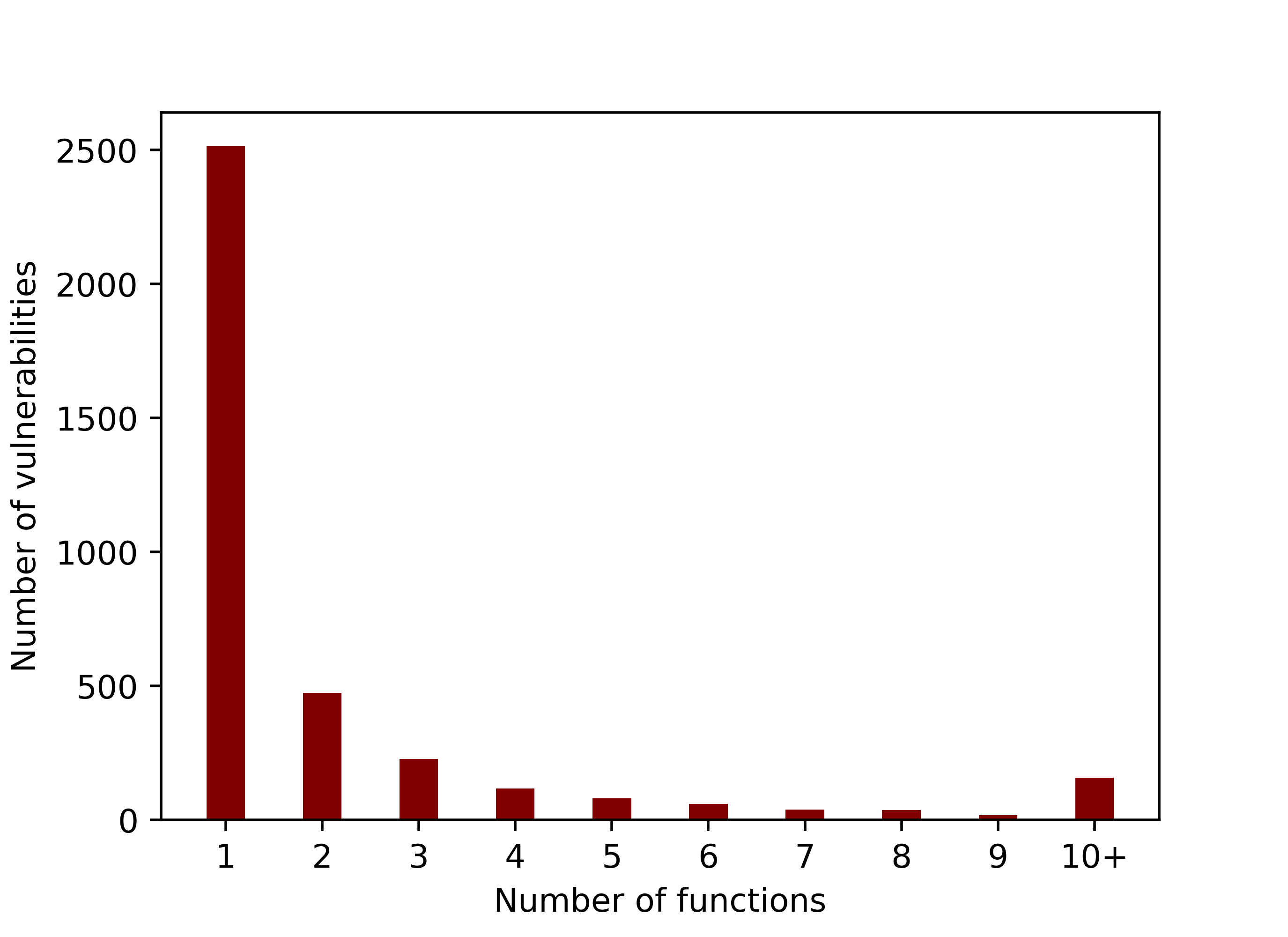}
    \caption{Vulnerabilities in the LineVul dataset, grouped by their number of functions }
    \label{fig:linevul-func-dist}
\end{figure}

In terms of functions, the LineVul dataset contained \lineVulFunctionComplexPatches{} function-based compound patches that fixed vulnerabilities, out of which \lineVulFunctionRepeatedVul{} were automatically found to be repeated IBU vulnerabilities. Removing these repeated IBU vulnerabilities from the compound patches revealed that 1212 or \lineVulFunctionComplexVul{}\% of the vulnerabilities in the LineVul dataset were actually function-based MBU vulnerabilities. The vulnerable functions in these MBU vulnerabilities accounted for \lineVulFunctionComplexContribution{}\% of all the vulnerable functions in the dataset. The distribution of functions per vulnerability in the LineVuln dataset can be found in Figure~\ref{fig:linevul-func-dist}. 

These results revealed that MBU vulnerabilities do form part of the datasets of these DL-based detectors and their base units contribute a significant chunk of the overall vulnerable base units. This further reinforced our argument that MBU vulnerabilities need to be taken into account when training and testing these models. 

\subsection{RQ2: Usage of MBU vulnerabilities in training and evaluation}
\label{sec:RQ2}
Knowing that MBU vulnerabilities contribute significantly to the datasets of our three subject DL-based vulnerability detectors, next, we set out to understand how are they used in learning (training) and evaluation (validation, testing, and accuracy reports). To do so, we looked at both the underlying approaches as well as the actual open-source implementations of the three detectors.

Through this part of the study, we aimed to ascertain whether the detectors ensured that (1) every vulnerability, along with all its base units, is always used either for training or testing (or validation, when applicable), exclusively,  and (2) when reporting accuracies, all of the base units of MBU vulnerabilities were accounted for. 

First, we explored the strategies the detectors used to assign each sample in their datasets into the traditional data splits associated with DL. Using MBU vulnerabilities either in training, validation, or testing, exclusively, is important to follow a realistic learning and evaluation scenario. The datasets of the three detectors were obtained through publicly available information. Various systems report their vulnerabilities either through the National Vulnerability Database (NVD) or their own websites. When MBU vulnerabilities are reported, all of their comprising base units are part of the same report simultaneously. The purpose of splitting the data into training, validation, and testing is to simulate a scenario in which the labels of the samples in training are known beforehand (historical data), whereas those of samples in validation and testing are unknown. To ensure that the DL model that is trained, validated, and tested on that data is realistic it is necessary to ensure that the assumption of which labels are known and which are unknown at training time is also realistic. Using a part of the same vulnerability for training and another part for testing, for example, does not simulate a realistic scenario, may taint the results, and thus may lead to incorrect conclusions. The importance of simulating a realistic scenario in DL applications has been brought up in previous work, where it was referred to as the ``realistic labeling assumption''~\cite{jimenez2019}. In their paper, Jimenez et al. discuss the importance of taking into account temporal constraints when using vulnerability data for training and testing~\cite{jimenez2019}. Breaking down MBU vulnerabilities into their base units and using them across training, validation, and testing also violates these constraints: it is incorrect to assume you know the labels of some of the base units involved in an MBU vulnerability (i.e., the ones that are assigned to the training set) and not others (i.e., the ones that are assigned to the validation or testing sets) when they are all reported and made publicly available at the same time. 

Our findings reveal that across all three detectors, MBU vulnerabilities are broken down across training and testing, and in cases when a validation set is used, even in validation. In fact, across all three approaches, we found an overlap between the train and test sets. ReVeal does not have a fixed training and testing set, so we used their code to generate train and test sets for models. We followed their default configuration for train and test separation across 30 runs. In those 30 runs, we saw that around 95\% of the MBU vulnerabilities had base units in both training and testing sets. In DeepWukong, 22\% of the MBU vulnerabilities had their base units broken down along the training, testing, and validation sets. Lastly, 36\% of LineVul's MBU vulnerabilities were included in the training, testing, and validation sets simultaneously. This suggests that these models are not trained, validated, and tested in a realistic way~\cite{jimenez2019}. We explore further the effects of learning and evaluating the three detectors in this form in Section~\ref{sec:RQ4}.

Second, we focused on how the three detectors include MBU vulnerabilities in their accuracy reports, which are part of their evaluation. Proper reporting of accuracies for vulnerability detection is crucial because it illustrates how useful these detectors can be in the real world. Accuracy reports where an MBU vulnerability's base units are not looked at comprehensively and in total are misguiding. When developers tackle a vulnerability, they need to know all of the comprising parts of the vulnerability; this is true for both MBU and IBU vulnerabilities. Approaches and implementations that claim to help developers find vulnerabilities need to ensure that their accuracy calculations reflect how well these approaches can find all of the vulnerable code locations. 

We found that all of the detectors fail to properly take into account MBU vulnerabilities in their accuracy reports: the MBU vulnerabilities are broken down into their comprising base units and the reported accuracies are per base unit. This stands in contrast to the claims in all of the three papers that the proposed approaches ``detect vulnerabilities,'' and not just their base units. For instance, the authors of ReVeal claim that ReVeal ``can find a larger number of true-positive vulnerabilities,'' when in reality they report accuracies on vulnerable functions and not complete vulnerabilities~\cite{chakraborty2021}. DeepWukong's first research question reads ``Can DeepWukong accurately detect vulnerabilities?'' but in their answer, the authors say they looked at individual slices, not complete vulnerabilities~\cite{cheng2021}. LineVul's first research question is similar: the authors ask how accurate LineVul is on ``function-level vulnerability predictions.'' However, later in the paper, they talk about predicting only vulnerable functions and not all of a vulnerability~\cite{fu2022}. It is important for these and other detectors to properly include MBU vulnerabilities in their accuracy calculations. 

\subsection{RQ3: Actual accuracies on MBU vulnerabilities}
\label{sec:RQ3}
After the results from the previous section, we set out to understand how the detectors' overall accuracy metrics are adjusted when taking into account complete vulnerabilities and also specifically how accurate the detectors are on MBU vulnerabilities. For the former, we looked at the precision, TPR (equivalent to recall), and the MCC metrics. For the latter, since we were specifically interested in how these tools detect vulnerabilities, i.e., the positive class, we wanted to isolate their accuracy only on vulnerabilities. That is why we looked at the TPR difference between IBUs and MBUs. 


Before we present the results of the three detectors, we will briefly present the data and the settings we used. 
 \begin{figure}[b!]
 \vspace{-1cm}
    \includegraphics[width=7cm]{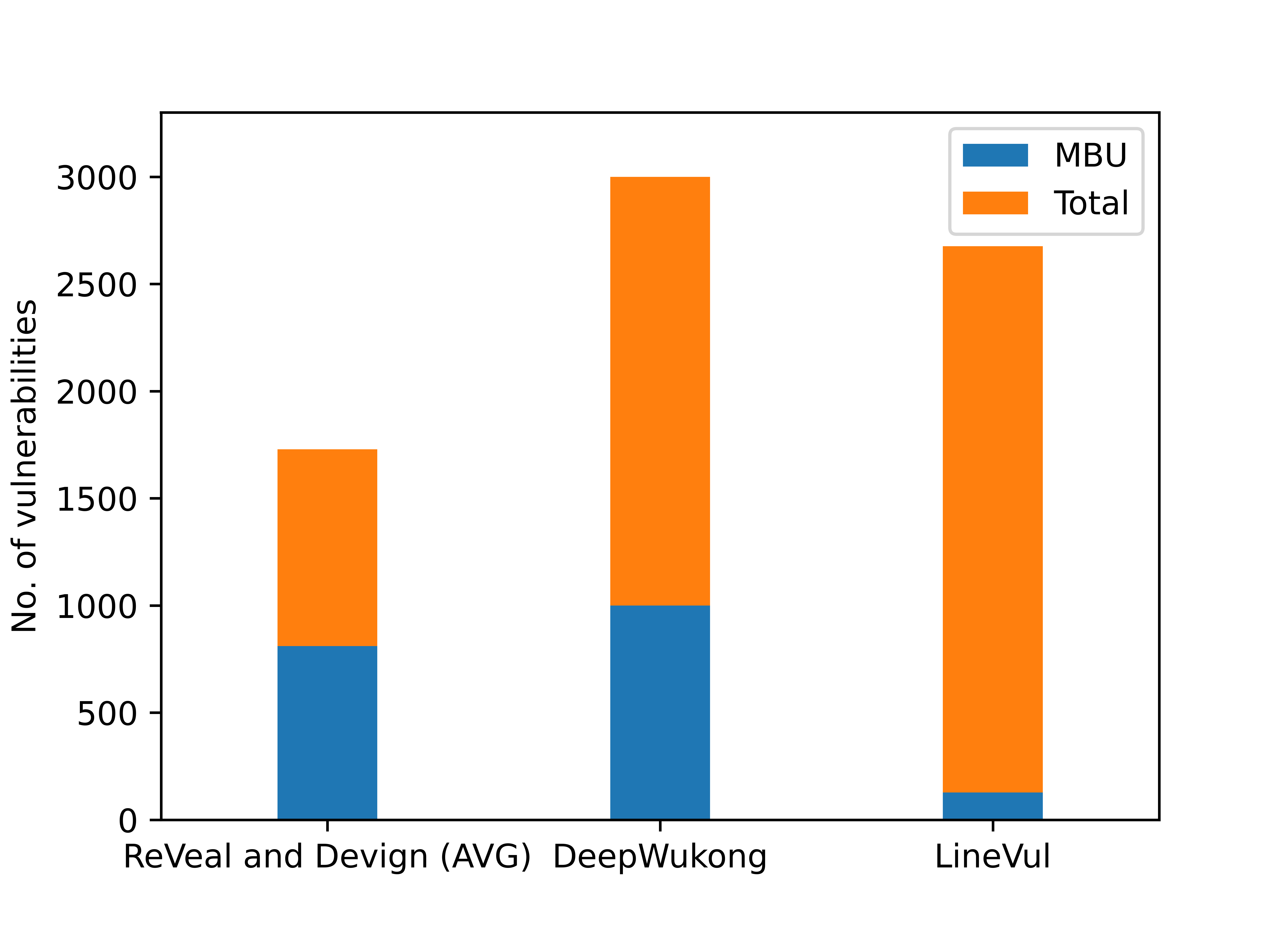}
    \caption{Vulnerabilities in the test sets of ReVeal (averages across the 30 runs, across which there was little variance), DeepWukong, and LineVul}
    \label{fig:dwk-linevul-ts-dist}
\end{figure}

ReVeal did not have a publicly available model at the time of our study, so we retrained their model with their publicly available code and with the portion of data for which we could obtain the relevant metadata (refer to Section~\ref{sec:prerq}). We followed their train, test, and iteration configurations. ReVeal by default does 30 iterations. DeepWukong and LineVul released their respective models as well as their fixed test sets, so we were able to use their trained models on their test sets.

The composition of the respective test sets for the three detectors can be found in Figure~\ref{fig:dwk-linevul-ts-dist}. For ReVeal and Devign, we present the averages of the 35 runs since we did not notice a lot of variance: around 1,729 vulnerabilities, out of which 811 are MBU. DeepWukong had around 1,000 MBU vulnerabilities out of the 3,000 in its test set, whereas LineVul had only 128 MBU vulnerabilities out of approximately 2,700 total vulnerabilities. 

A final note on LineVul's test set: LineVul assigns vulnerability scores to lines in functions it identifies as vulnerable. It then reports the Top-10 accuracy for localizing vulnerable lines per function, i.e., the percentage of functions for which at least one actual vulnerable line is included in the top 10 lines ranked by their vulnerability score. Because of the score, we were unable to obtain fine-grained results about which specific lines LineVul predicts as vulnerable; we need the fine-grained results per each line for our subsequent analysis of accuracies on complete vulnerabilities. The lack of fine-grained results rendered getting the accuracies for the line-based setting of LineVul impossible. Thus, we present the results only at the function level. 
\begin{table}[]
\begin{tabular}{|l|l|cc|c|c|}
\hline
                               &                   & \multicolumn{2}{c|}{\textbf{ReVeal}}                   & \multirow{2}{*}{\textbf{DWK}} & \multirow{2}{*}{\textbf{LV}} \\ \cline{1-4}
\textbf{Metric}                & \textbf{Setting}  & \multicolumn{1}{c|}{\textbf{Mean (SD)}} & \textbf{Med} &                               &                              \\ \hline
\multirow{4}{*}{\textbf{TPR}}  & \textbf{Base}     & \multicolumn{1}{c|}{0.91 (0.04)}        & 0.91         & (0.93)                        & \lineVulFuncLevelTPRFunction \\ \cline{2-6} 
                               & \textbf{Adjusted} & \multicolumn{1}{c|}{0.90 (0.05)}        & 0.89         & (0.92)                        & \lineVulFuncLevelTPRVul      \\ \cline{2-6} 
                               & \textbf{IBU}      & \multicolumn{1}{c|}{0.92 (0.04)}        & 0.92         & (0.92)                        & \lineVulFuncLevelTPRAtomic   \\ \cline{2-6} 
                               & \textbf{MBU}      & \multicolumn{1}{c|}{0.77 (0.09)}        & 0.76         & (0.91)                        & \lineVulFuncLevelTPRComplex  \\ \hline
\multirow{2}{*}{\textbf{Prec}} & \textbf{Base}     & \multicolumn{1}{c|}{0.49 (0.01)}        & 0.48         & (0.94)                        & 0.99                         \\ \cline{2-6} 
                               & \textbf{Adjusted} & \multicolumn{1}{c|}{0.41 (0.01)}        & 0.41         & (0.94)                        & 0.99                         \\ \hline
\multirow{2}{*}{\textbf{MCC}}  & \textbf{Base}     & \multicolumn{1}{c|}{0.13 (0.04)}        & 0.14         & (0.92)                        & 0.91                         \\ \cline{2-6} 
                               & \textbf{Adjusted} & \multicolumn{1}{c|}{0.11 (0.04)}        & 0.11         & (0.92)                        & 0.90                         \\ \hline
\end{tabular}
\vspace{0.2cm}
\caption{TPR, Precision (Prec), and MCC, of ReVeal, DeepWukong (DWK), and LineVul (LV) in their test sets, per base unit, i.e., their original way of measuring accuracy (\textit{Base}), per vulnerability (\textit{Adjusted}), and TPR per IBU vulnerabilities (\textit{IBU}) and per MBU vulnerabilities (\textit{MBU}). For ReVeal, we present the mean, standard deviation (SD), and median (Med) for the 30 runs.}
\label{tab:overview-results}
\vspace{-1cm}
\end{table}

The overview of the results from this part of our study can be found in Table~\ref{tab:overview-results}. For each of the detectors, we first calculated their accuracy metrics per vulnerability base unit (the \textbf{Base} rows), as they do in their original approaches and implementations. In this case, a \textit{true positive} instance is any vulnerable base unit correctly predicted, whereas a \textit{false negative} instance is any vulnerable base unit labeled as non-vulnerable. We then adjust the original metric (results presented in the \textbf{Adjusted} rows) by considering vulnerabilities, i.e., grouping all base units of an MBU vulnerability, and defining a \textit{true positive} instance only if all of the base units of a vulnerability have been correctly predicted (row \textbf{Adjusted} in the table). In the same vein, a \textit{false negative} is defined as those vulnerabilities for which the detector fails to identify all base units as vulnerable. The definitions of false positive and true negative remain unchanged. 

For ReVeal, for each metric, we report the average along with standard deviation (column \textbf{Mean (SD)}) and the median of the 30 iterations. The complete results from all 30 iterations can be found on our website~\cite{ourwebsite}. 

The DeepWukong results are within the parenthesis to indicate the fact that they are an upper bound since, as explained above, we assumed their multi-slice test cases perfectly correspond to MBU vulnerabilities.

In regards to the vulnerable samples, some of the results confirmed our expectations. All three detectors over-report their TPR, albeit DeepWukong and ReVeal do so by small margins only. LineVul over-reports them by 3 percentage points. However, the TPRs on MBU vulnerabilities drop significantly by 15-26 percentage points. Since these detectors do not account for MBU vulnerabilities when training, it is expected to see this type of performance. 

We note DeepWukong's very high accuracy upper bound across the board. Unfortunately, we do not have access to the raw data used in their datasets which limited our ability to further inspect the results. DeepWukong has the highest percentage of MBU vulnerabilities in its dataset, which may indicate there is a correlation between including a higher rate of MBU vulnerabilities and higher TPRs. Whether including more MBU vulnerabilities leads to better results merits further exploration and research. We should note that in studies with the other two datasets, DeepWukong's model was not able to replicate the same levels of accuracy; in fact, the adjusted accuracy metrics and the TPR on MBUs from these studies follow the same trend as the metrics of ReVeal and LineVul seen in Table~\ref{tab:overview-results}.

The results on the precision and MCC metric offer another perspective on the performance of the three detectors. ReVeal's precision drops the most when considering complete vulnerabilities. LineVul's metrics decrease slightly when we use our adjusted metrics. DeepWukong's upper bounds for the metrics remain unchanged in the two runs. 

The difference in how the metrics change between the Base and Adjusted cases illustrates that the vulnerability detectors have different strengths. Including all the metrics in the accuracy reports helps the consumers of these detectors to make more informed decisions about when to use them. 

\subsection{RQ4: Training and evaluating under a realistic scenario}
\label{sec:RQ4}

Our RQ2 also revealed that the three detectors do not take into account complete vulnerabilities when training, validating, and testing, splitting the base units of the same vulnerability across the three different sets. Our final research question explores the impact of this on the accuracies of the three detectors. We hypothesized that the accuracies are over-reported.

\begin{table}[]
\begin{tabular}{|l|l|cc|c|c|}
\hline
                               &                   & \multicolumn{2}{c|}{\textbf{ReVeal}}                   & \multirow{2}{*}{\textbf{DWK}} & \multirow{2}{*}{\textbf{LV}} \\ \cline{1-4}
\textbf{Metric}                & \textbf{Setting}  & \multicolumn{1}{c|}{\textbf{Mean (SD)}} & \textbf{Med} &                               &                              \\ \hline
\multirow{4}{*}{\textbf{TPR}}  & \textbf{Base}     & \multicolumn{1}{c|}{0.87 (0.04)}        & 0.87         & (0.93)                        & 0.91                         \\ \cline{2-6} 
                               & \textbf{Adjusted} & \multicolumn{1}{c|}{0.87 (0.04)}        & 0.88         & (0.90)                        & 0.86                         \\ \cline{2-6} 
                               & \textbf{IBU}      & \multicolumn{1}{c|}{0.91 (0.03)}        & 0.92         & (0.89)                        & 0.93                         \\ \cline{2-6} 
                               & \textbf{MBU}      & \multicolumn{1}{c|}{0.66 (0.08)}        & 0.67         & (0.91)                        & 0.73                         \\ \hline
\multirow{2}{*}{\textbf{Prec}} & \textbf{Base}     & \multicolumn{1}{c|}{0.51 (0.03)}        & 0.52         & (0.97)                        & 0.54                         \\ \cline{2-6} 
                               & \textbf{Adjusted} & \multicolumn{1}{c|}{0.40 (0.03)}        & 0.40         & (0.93)                        & 0.31                         \\ \hline
\multirow{2}{*}{\textbf{MCC}}  & \textbf{Base}     & \multicolumn{1}{c|}{0.08 (0.04)}        & 0.08         & (0.94)                        & 0.68                         \\ \cline{2-6} 
                               & \textbf{Adjusted} & \multicolumn{1}{c|}{0.11 (0.04)}        & 0.12         & (0.91)                        & 0.50                         \\ \hline
\end{tabular}
\vspace{0.2cm}
\caption{TPR, Precision (Prec), and MCC, of ReVeal, DeepWukong (DWK), and LineVul (LV) when trained, validated and tested focusing on vulnerabilities, per base unit, i.e., their original way of measuring accuracy (\textit{Base}), per vulnerability (\textit{Adjusted}), and TPR per IBU vulnerabilities (\textit{IBU}) and per MBU vulnerabilities (\textit{MBU}). For ReVeal, we present the mean, standard deviation (SD), and median (Med) for the 30 runs.} 
\vspace{-1cm}
\label{tab:constrained-training-testing}
\end{table}

This research question required retraining the three detectors. We followed the original approaches to retrain, revalidate, and retest the detectors in every way but one: instead of focusing on base units
when splitting the data, we focused on complete vulnerabilities. We present the results of this experiment in Table~\ref{tab:constrained-training-testing}.  \begin{wrapfigure}[19]{r}{0.2\columnwidth}
\vspace{-1.2cm}
  \hspace*{-0.8cm}
    \centering
    \includegraphics[width=12cm]{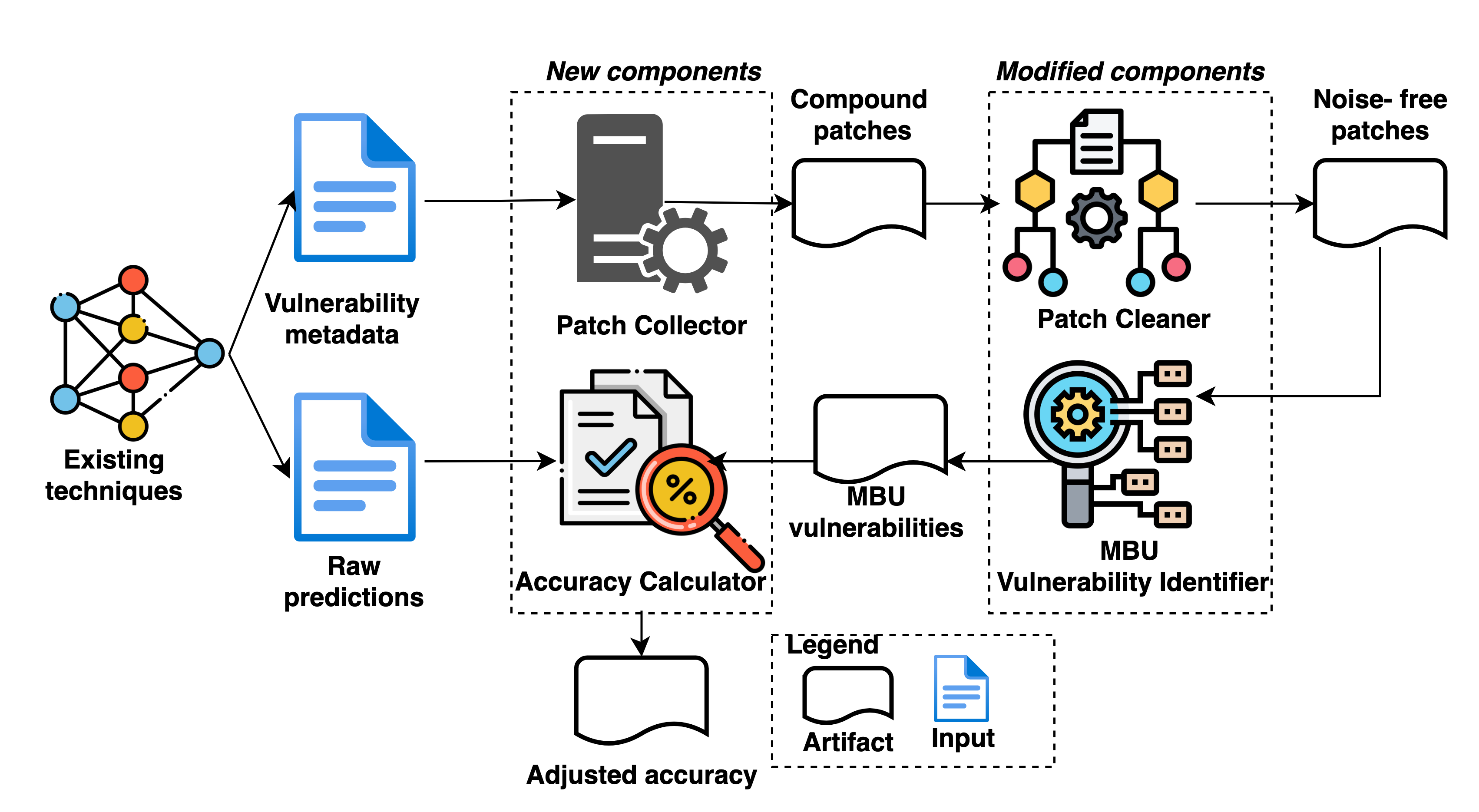}
\captionsetup{width=12cm,justification=centering}  
    \caption[Framework for detecting and understanding MBU vulnerabilities]{Framework for detecting and understanding MBU vulnerabilities}
    \label{fig:framework}

%
\end{wrapfigure}

These results pointed to interesting underlying phenomena, at times confirming and at others rejecting our hypothesis. To make sense of these results, we compare them to those presented in Table~\ref{tab:overview-results}, depicting the original results for the three detectors. ReVeal's 
performance across most metrics (with the exception of precision) suffers when trained and tested realistically. This matches our expectations and hypothesis. However, LineVul's performance on the TPR metrics mostly improves under realistic training! On the other hand, its performance on precision and MCC drops significantly. One explanation for the contradicting changes in LineVul's metrics is that LineVul balances its training set: originally, there were as many vulnerable functions as there were non-vulnerable functions. Under the realistic training constraint, we balanced the number of complete vulnerabilities to non-vulnerable samples. This meant that the number of vulnerable functions in the training set was higher than in the original training of LineVul, which corresponded to better performance in detecting vulnerabilities, but worse performance in detecting non-vulnerable samples. The trade-off between precision and recall in models like LineVul's merits further exploration. DeepWukong's results should again be interpreted as an upper bound on its performance. They do not change significantly under the realistic training constraint. 
\vspace{-0.1cm}
\section{Framework}
\label{sec:framework}


We release a framework consisting of components we used in this study. We aim to facilitate the process of studying, understanding, and detecting MBU vulnerabilities. The framework can be used towards improving the way DL-based detection techniques handle MBU vulnerabilities, but also for further exploration of different aspects of the current DL-based vulnerability detectors and their datasets. An overview of our framework, its components, and how it can be used is depicted in Figure~\ref{fig:framework}. We explain the components and the usage we envision for our framework next.

\textit{Patch Collector} obtains vulnerability patches used in datasets. It takes as input vulnerability metadata, i.e., patch hashes and repository names where such patches can be obtained as well as the base unit the vulnerabilities are expressed in. Optional metadata that can be included
as part of the input is a count of the base units. If this optional metadata is given, the Patch Collector obtains only the compound 
patches directly. Otherwise, it first obtains the code changes in the patch, groups them per base unit, and then concludes whether the patch should be kept (if it is compound) for further analysis. Patch Collector, then, represents the changes that have happened in the patch in the form of AST edit scripts, by using GumTree~\cite{falleri2014}. 

We have also added a \textit{Patch Cleaner} component. As mentioned above, previous research studies have complained about the presence of noise in vulnerabilities~\cite{chakraborty2021, sejfia2021}. \vspace{17\baselineskip}  At the same time, some \\ \\ \\ \\ studies show how to detect part of that noise~\cite{kawrykow2011, sejfia2021}. To facilitate the process of removing noise, we have included the CasCADe technique~\cite{sejfia2021}, a rule-based noise-detection technique, in our Patch Cleaner component. 

We did not originally include the Patch Cleaner in our study presented in Section~\ref{sec:findings} because Patch Cleaner would alter the ground truth. If the Patch Cleaner found noise in the data, we would have removed that noise; the ground truth collected by the detectors would have changed, which would not result in an accurate comparison between the reported accuracy metrics in the original papers and the accuracy metrics we obtained. To demonstrate the usage of the Patch Cleaner we did, however, run an experiment on a portion of the data used by ReVeal, around 1000 compound vulnerability-fixing patches. Since ReVeal uses functions as the base unit, using CasCADe, we identified functions that contained only API-based casualty changes or logic-preserving changes that happen inside functions because of changes to APIs. These changes are considered as not pertaining to the vulnerability, thus the functions that solely contain them should not be considered vulnerable. We found 25 functions, spread around 35 patches, had solely casualty changes. The low occurrence of casualty changes was expected in this case because the dataset was manually labeled~\cite{zhou2019} and thus less likely to contain noise. 

Our Patch Cleaner, following CasCADe's initial approach, is rule-based and thus can be extended. New rules or techniques on how to detect noise can be added to it. They would require a one-off integration with the rest of the framework.  

\textit{MBU Vulnerability Identifier} takes the AST-represented patches as input and identifies which ones are MBU vulnerabilities. It follows the similarity-based approach introduced in Section~\ref{sec:identification}, which formed part of C3's approach~\cite{kreutzer2016}. The Identifier can be configured to look for changes based on the various base units. Further, following C3's original approach it encodes the AST-based edit scripts for faster processing. This component uses the longest-common-subsequence method to calculate the similarity. 

Finally, using the identified MBU vulnerabilities and raw predictions, the \textit{Accuracy Calculator} provides accuracy metrics per \textit{vulnerability}. The metrics can be further expanded. 

We envision our framework to be used in tandem with the development of new DL-based vulnerability detectors and perhaps even to improve existing ones. For instance, the output of our MBU Vulnerability Identifier can be used early in the data collection stage to ensure that enough MBU vulnerabilities are present in the dataset. Further, the output of this component should be tightly coupled with the selection of training and testing sets. As we found out, DL-based detectors do not currently ensure MBU vulnerabilities are used exclusively either for training or testing. With our framework, they can obtain information that helps them realistically set their training and testing processes. In addition, with the help of Patch Cleaner, DL-based detectors can be more confident in the quality of their data. Finally, we argue that DL-based detectors should report our Accuracy Calculator metrics. It is crucial to show accuracy metrics per vulnerability to more realistically represent the usefulness of these detectors. But it is also important to specifically report accuracy metrics for MBU vulnerabilities. These vulnerabilities, precisely because they span multiple code locations that have intricate dependencies, may be more challenging for developers to find on their own, even more so than IBU ones. 

Further, other researchers can reuse our framework to analyze other DL-based vulnerability detectors. Detecting vulnerabilities using DL-based solutions seems to be a growing field. But at the same time, it is important to verify these solutions are useful for real-world scenarios. Our framework helps with that. Other researchers can also expand our framework. One important way in which that can be done is through expanding the types of base units we consider. Vulnerability detection tools, for instance, at times focus on coarser-grained components; in the future, other types of representations (i.e., data-flow only, AST-based, or specific code regions not bound within the current base units) may become prevalent. Supporting a variety of types of base units helps facilitate the process of verifying the results and usefulness of DL-based approaches. 

\vspace{-0.2cm}
\section{Related Work}
\label{sec:related-work}

DL-based detectors are the main subject of this study. The three detectors we chose to analyze represent examples of prominent vulnerability detection approaches. Others include Devign, a function-level DL-based vulnerability detector~\cite{zhou2019}. The portion of the ReVeal dataset we used in this study originally came from the authors of Devign. VulDeePecker~\cite{li2018} and Sysevr~\cite{li2021} propose a combination of slice-based detection of vulnerabilities. Finally, LineVD detects vulnerabilities at the line level~\cite{hin2022}. These DL-based studies have been enabled by vulnerability data collection efforts such as the work of Fan et al.~\cite{fan2020}, which was used by LineVul and many other research studies. 

While pattern-based detectors are being outnumbered by DL-based ones, initial studies on how to best represent code for vulnerabilities used patterns~\cite{yamaguchi2014} and have inspired better representations for code even in techniques used today. While in this study we looked at specifically DL-based detectors, the concept of MBU vulnerability applies to other detectors as well. We can test pattern-based detectors on how well they detect MBU vulnerabilities. 

The accuracy of the DL and other data-based detectors has been demonstrated to vary quite a lot when the data changes~\cite{chakraborty2021, jimenez2019}. We have also noticed this in our own experiments. Because of this, the quality of the data included in vulnerability datasets has been the subject of previous research~\cite{croft2023data}. Security data quality transcends vulnerability detection: security bug report prediction has also gained traction and data quality matters for this type of insight too~\cite{wu2021data, zheng2022domain}. Our study, in a way, also speaks to the underlying quality of the data used in vulnerability detection: we explore if the datasets contain MBU vulnerabilities. In addition, we present a new way to approach datasets that contain them. 

Croft et al. in their work compile the processes of data preparation for vulnerability detection~\cite{croft2022data}. In fact, they do refer to the usage of base units and how that constrains vulnerability detection approaches, though they do not refer to a base unit by that name. With the boom of data-driven vulnerability detection, understanding vulnerability datasets has also become important~\cite{li2023anatomy}. Our study also provides a deeper understanding of vulnerability databases through analyzing their inclusion of MBU vulnerabilities. 

Other aspects of vulnerability datasets have also been studied. For instance, Le et al. check how vulnerability data can be used to prioritize vulnerabilities~\cite{le2022survey}. Proper prioritization of vulnerabilities has also been studied by looking at the inconsistencies that can exist in vulnerability severity scores~\cite{croft2022investigation}. The authors claim that severity scores are not properly considered when reporting vulnerabilities, affecting downstream tasks. While prioritization was not the main intention behind our work, our analysis of the data can be used to categorize vulnerabilities and prioritize them. For instance, detectors may rank MBU vulnerabilities higher than IBU ones. 

Looking at how vulnerability detectors perform in realistic scenarios has been the focus of the work by Jimenez et al.~\cite{jimenez2019}. ReVeal, besides the detection platform it provides, also studies other DL-based detectors in realistic scenarios~\cite{chakraborty2021}. Our work in identifying MBU vulnerabilities complements these existing efforts. 

Finally, we made use of clustering code changes in our work. Our approach was a modification of C3's approach~\cite{kreutzer2016}. Others have looked at how to mine patterns from changes~\cite{martinez2013automatically}. To the best of our knowledge, we are the first to apply code clustering to distinguish IBU from MBU vulnerabilities.
\section{Limitations and Threats to Validity}
\label{sec:threats}
Parts of our work relied on manual steps and third-party tools, both of which could introduce errors in our study and threaten its validity. We discuss the steps we took to mitigate these potential errors. 

1\textit{) Internal Validity:} We manually collected two ground truth datasets in this study to determine the accuracy of our MBU Vulnerability Identifier. The manual collection of the data exposed us to potential biases being introduced. However, to mitigate biases and errors, the original ground truth dataset involved three experts with several years of experience. One person worked on the manual tagging of the verification dataset but we release both the ground truth datasets publicly so that the research community can verify our claims as well. 

We rely on several third-party tools for our collection and analysis of data. For instance, we use GumTree for AST-based edit script generation and we are bound by its limitations. When possible, we have attempted to boost its accuracy by improving the alignment of the trees. We use LLVM~\cite{llvm} in the Patch Cleaner and are bound by the limitations of static analysis. 
 
2)\textit{ External Validity:} Our usage of the three particular datasets and three detectors is an external validity threat. We focus on three DL-based detectors and their dataset in this study. The generalizability of our conclusions because of this is also a threat to our study's validity. To mitigate this threat, we picked detectors with different base units, illustrating the adaptability of the concept of MBU vulnerabilities. 

We aimed to reproduce the results of the three detectors as faithfully as possible, to ensure we can properly illustrate their performance on the MBU vulnerabilities defined in their own datasets. However, we were also unable to obtain part of the data used in ReVeal and DeepWukong due to the lack of metadata. Moreover, our usage of these three detectors renders the study subject to the limitations and biases of the detectors. It should be noted, though, that our purpose was to realistically establish these detectors' performance, and their limitations and biases impact that performance. 

Further, all the vulnerabilities we analyzed exist in code written in C/C++. Further studies are needed to analyze the presence of MBU vulnerabilities in other programming languages. We expect our study to facilitate this exploration since the concepts used in the paper are not limited to any programming language. Our proposed framework, which can be directly used in such exploration, is language agnostic as well. 

3)\textit{ Construct Validity:} Accurate distinguishing of compound patches is necessary to mitigate construct validity threats. We have based both our definition and automated approach on real-world patches and real-world vulnerabilities. Our approach has been shown to be effective and we make the results publicly available, in order to enable further inspection.

\section{Conclusion and Future Work}
\label{sec:conclusion}
In this study, we introduce the concept of MBU vulnerabilities. Our study of three prominent DL-based detectors, ReVeal, DeepWukong, and LineVul, suggests that currently these detectors are not trained and tested in a realistic fashion, and further, they do not report accuracies on complete vulnerabilities, failing to include all components of MBU vulnerabilities. Our goal through this study was to establish the practices of these DL-detectors in including MBU vulnerabilities across their learning and evaluation process as well as their abilities to properly detect such vulnerabilities.

Our study indicates that existing DL-based detectors do not work as well in detecting all comprising parts of MBUs. This matched our expectations as these DL-based detectors do not seem to work well in data they are not trained on. This performance sheds light on the effectiveness of these detectors for detecting vulnerabilities as a whole, especially when these vulnerabilities are MBUs.

The message of this study for researchers of DL-based vulnerabilities is to (1) change how training and testing are done so that these processes can better reflect a realistic scenario and (2) include all of the comprising parts of a vulnerability in their accuracy metrics. We have grouped components from our study into a framework that helps to that end. 

Our future plans in this area are two-fold. First, we aim to further investigate the nature of MBU vulnerabilities. We aim to establish why certain vulnerabilities are spread along multiple locations and whether such vulnerabilities share some characteristics. Our intuition is that characteristics such as their severity or root cause may impact the spread of MBU vulnerabilities along their base units. Second, we plan to further inspect the correlation between the mere presence of MBU vulnerabilities in the training set and improved vulnerability detection. Our goal is to conduct a systematic study to properly establish the factors that impact the performance of the detectors on MBU vulnerabilities. 


\begin{acks}
This work is supported by a Google PhD Fellowship, the Department of Homeland Security under award number 70RCSA22C00000008, subaward number A2774-03, the National Science Foundation under award number 182335, and the Austrian Science Fund under grant number FWF P31989-N31.
\end{acks}
\balance
\bibliographystyle{ACM-Reference-Format}
\bibliography{acmbibl.bib}
\end{document}